# µW-Level Microresonator Solitons with Extended Stability Range Using an Auxiliary Laser

SHUANGYOU ZHANG[1], JONATHAN SILVER[1], LEONARDO DEL BINO[1,2], FRANCOIS COPIE[1], MICHAEL T. M. WOODLEY[1,2], GEORGE N. GHALANOS[1,3], ANDREAS SVELA[1,3], NIALL MORONEY[1,3], AND PASCAL DEL'HAYE[1*]

[1]*National Physical Laboratory (NPL), Teddington TW11 0LW, United Kingdom*
[2]*Heriot-Watt University, Edinburgh, EH14 4AS, Scotland*
[3]*Imperial College London, SW7 2AZ, United Kingdom*
*Corresponding author: pascal.delhaye@npl.co.uk*

**The recent demonstration of dissipative Kerr solitons in microresonators has opened a new pathway for the generation of ultrashort pulses and low-noise frequency combs with gigahertz to terahertz repetition rates, enabling applications in frequency metrology, astronomy, optical coherent communications, and laser-based ranging. A main challenge for soliton generation, in particular in ultra-high-Q resonators, is the sudden change of circulating intracavity power during the onset of soliton generation. This sudden power change requires precise control of the seed laser frequency and power or fast control of the resonator temperature. Here, we report a robust and simple way to increase the stability range of the soliton regime by using an auxiliary laser that passively stabilizes the intracavity power. In our experiments with fused silica resonators, we are able to extend the pump laser frequency stability range of microresonator solitons by two orders of magnitude, which enables soliton generation by slow and manual tuning of the pump laser into resonance and at unprecedented low power levels. Both single- and multi-soliton mode-locked states are generated in a 1.3-mm-diameter fused silica microrod resonator with a free spectral range of ~50.6 GHz, at a 1554 nm pump wavelength at threshold powers <3 mW. Moreover, with a smaller 230-µm-diameter microrod, we demonstrate soliton generation at 780 µW threshold power. The passive enhancement of the stability range of microresonator solitons paves the way for robust and low threshold microcomb systems with substantially relaxed stability requirements for the pump laser source. In addition, this method could be useful in a wider range of microresonator applications that require reduced sensitivity to external perturbations.**

## 1. INTRODUCTION

Over the past two decades, optical frequency combs based on mode-locked lasers have revolutionized the field of precision spectroscopy with an unprecedented frequency measurement precision [1, 2]. They have attracted great attention in many areas, such as optical clocks [3, 4], ultralow-noise microwave generation [5, 6], photonic ADCs and radar [7, 8], and gas monitoring [9], just to name a few. Nowadays, Kerr frequency combs ("microcombs") based on parametric four-wave mixing in monolithic high-Q microresonators provide an alternative scheme to miniaturize comb systems and enable out-of-lab applications [10-12]. After their first demonstration, microcombs were soon demonstrated in resonators made from silica [10, 13], magnesium fluoride ($MgF_2$) [14, 15], silicon nitride ($Si_3N_4$) [16-21], silicon (Si) [22], diamond [23], hydex [24], and several other materials. More recently, dissipative Kerr soliton formation has been observed in microresonators providing low noise, fully coherent Kerr frequency combs with ultrashort optical pulse trains and reproducible spectral envelopes [14] spanning up to more than an octave [21, 25]. So far, *soliton* microcombs have been reported in $MgF_2$ [14, 15], $Si_3N_4$ [18-21], silica [26, 27] and Si [28] with repetition rates ranging from 1 THz [21, 25] down to 1.8 GHz [29], at wavelengths all the way from the visible [30] to the mid-infrared [28]. Indeed, soliton microcombs have already been successfully used for optical frequency synthesizers [31], astronomy [32, 33], optical coherent communications [34], laser-based light detection and ranging [35, 36], and dual-comb spectroscopy [37-39]. Moreover, soliton crystals [40] and Raman Stokes solitons [41] have been recently observed in microresonator systems.

The formation of a single soliton in microresonators typically requires the pump frequency to be red-detuned relative to the thermally-shifted cavity resonance [14]. However, due to the thermal and Kerr response of the cavity modes to laser fluctuations, red-detuned pump frequencies are unstable while blue-detuned frequencies are stable [42]. As a result, accessing soliton states is experimentally challenging. To trigger the soliton states, different methods have been developed, such as power kicking [43, 44], frequency kicking [45, 46], and thermal control [47]. Direct soliton generation can also be achieved by optimizing laser tuning speed and stopping at the right frequency. This method works in materials with weak thermo-optic effect like $MgF_2$ [14]. Power and frequency kicking methods are based on abrupt changes in the pump power or frequency, respectively. These changes are much faster than the thermal drift of the resonator modes. Recent work in $Si_3N_4$ and $MgF_2$ microresonators demonstrated that multi-soliton states can be deterministically switched to single-soliton states by reducing the number of solitons one by one through backward tuning of the pump frequency [20]. In addition, very recently, it has been demonstrated that spatial mode-interaction in microresonators can support soliton generation [21, 48].

A key challenge for most applications using microresonator-based frequency combs is the stable long term operation of the comb. In this work, we report the passive enhancement of the stability range of microresonator solitons by using an auxiliary laser. The auxiliary laser passively compensates thermal fluctuations and Kerr shifts of the resonator modes due to drifts and fluctuations of the soliton pump laser. In addition, the auxiliary laser compensates sudden intracavity power changes when the microresonator enters the soliton regime. Using this method, the length of the soliton stability range (soliton steps) is extended from 100 kHz to 10 MHz, which enables access to single-soliton states without specific requirements for pump laser tuning speed or power kicking techniques. Soliton states can be reached by arbitrary slow tuning of the laser into resonance, which significantly simplifies the soliton generation process. In particular this enables the access to soliton states in ultra-high-Q resonators with flawless mode spectra (no mode crossings), which has been previously challenging. The enhanced stability range enables us to generate solitons at a very low threshold power of 780 µW in a 230-µm-diameter microrod resonator (280 GHz mode spacing). In addition we demonstrate single- and multi-soliton states at 3 mW threshold power in 1.3-mm-diameter glass rods (50 GHz mode spacing). The single-soliton optical spectrum has a smooth, $sech^2$-like shape without significant imperfections due to mode-crossings. Low power consumption of microresonator solitons is in particular important for out-of-the-lab applications of frequency combs e.g. in battery powered systems [49].

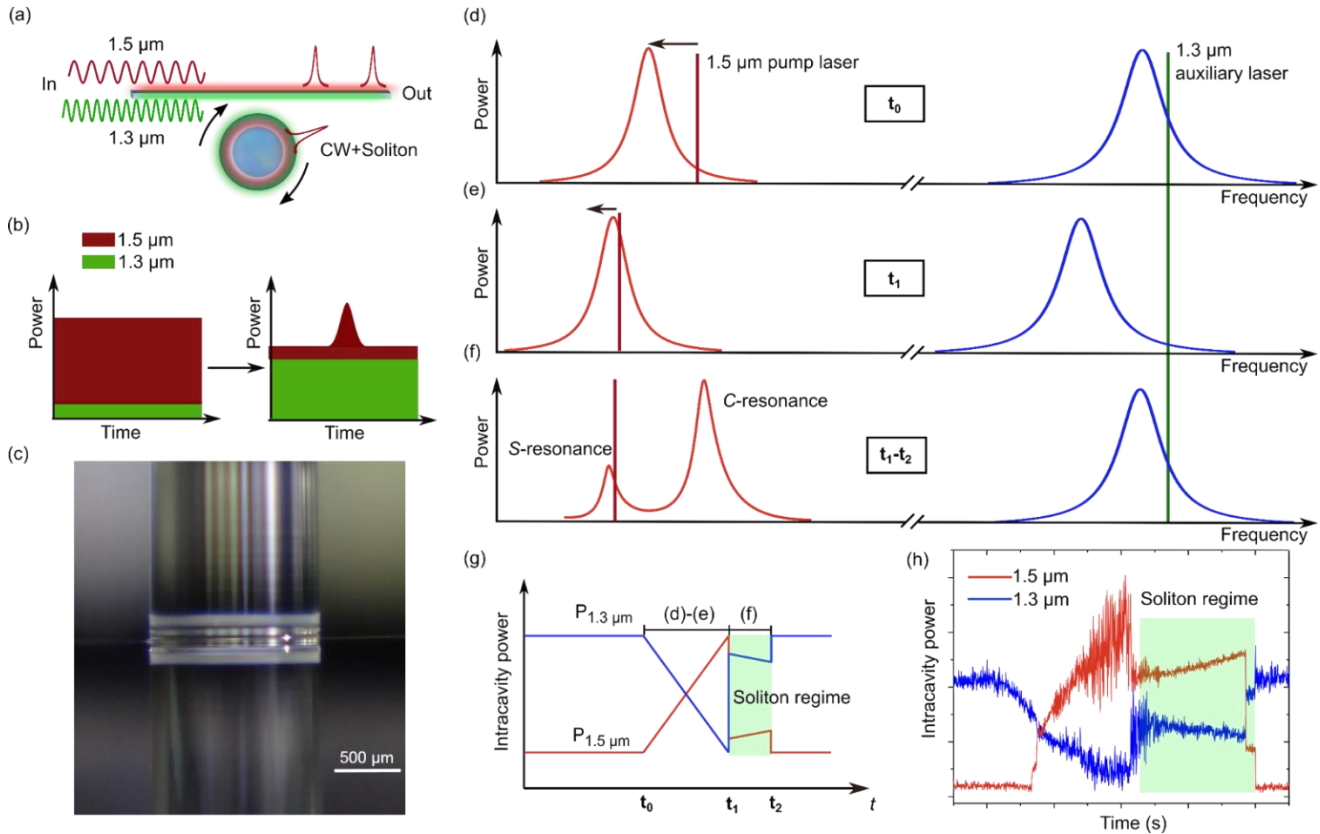

**Fig. 1.** (a) Scheme of using an auxiliary laser to assist in accessing soliton states. The 1.5-µm-pump-laser generates a microresonator soliton while the second laser at 1.3 µm wavelength passively stabilizes the intracavity power. (b) Illustration of the intracavity power (red: 1.5 µm, green: 1.3 µm) before (left panel) and after (right panel) soliton generation. (c) Microscope image of the 1.3-mm-diameter fused silica microrod resonator used in the experiments. Panels (d)-(f) show the principle of the passive compensation of the circulating power in the microresonator by the auxiliary laser in order to enhance the soliton stability range (animated version available as supplemental material). In panel (d) only the auxiliary laser is coupled into a resonator mode. When tuning the pump laser into resonance, shown in panel (e), the thermal shift of the resonator modes automatically reduces the amount of light coupled into the auxiliary resonator mode. Panel (f) shows the abrupt transition into a soliton state, which reduces the coupled power of the pump laser. In this state, the pump resonance splits into a C-resonance (resonance for light arriving out-of-sync with the soliton) and an S-resonance (resonance for light arriving in-sync with the soliton). The reduction in pump power in the soliton regime moves the auxiliary resonance back towards the auxiliary laser and thus compensates the power loss. Panel (g) shows the temporal evolution of the intracavity power when tuning the pump laser into resonance with a fixed frequency auxiliary laser. Panel (h) shows an experimental trace according to the scheme in panel (g). The 1.3-µm-auxiliary laser passively compensates changes of circulating power of the pump laser.

## 2. PRINCIPLE OF PASSIVE SOLITON STABILIZATION

Figure 1(a) shows the concept of using an auxiliary continuous wave laser to increase the stability range for microresonator solitons. The auxiliary laser at 1.3 µm is kept at a fixed frequency on a resonator mode while a soliton is generated by a second laser at 1.5 µm. The 1.3 µm laser provides a background signal that compensates fluctuations of the 1.5-µm soliton laser at time scales that are slower than the cavity build-up time. At the output, the soliton pulse can be separated using a wavelength division multiplexer (WDM). Figure 1(b) illustrates the intracavity power before and after entering the soliton state. A small portion of the 1.3 µm auxiliary laser power is coupled into the resonator prior to the soliton generation. Once the soliton is formed, the intracavity power of the auxiliary laser passively rises to compensate the temperature variation of the resonator caused by the loss of intracavity pump power. In the soliton state, the intracavity field consists simultaneously of an intense soliton pulse, weak 1.5 μm CW background and 1.3 μm auxiliary CW background. The resonator used in the experiment is a fused silica microrod resonator, shown in Fig. 1(c) [50] with a Q-factor of $2\times10^8$ at 1.3 μm and $3.7\times10^8$ at 1.5 μm. By controlling the curvature of the resonator sidewalls during $CO_2$ laser machining, the microrod resonator can be engineered to have an ultrahigh optical quality factor and minimal avoided crossings between different mode families.

Figure 1(d) - (f) show how the auxiliary laser stabilizes the optically circulating power within the resonator to enhance the stability range of the soliton generation. First, the frequency of the 1.3 μm auxiliary laser is tuned into a high-Q optical resonator mode from the blue side and fixed on the blue side of the resonance, as shown in Fig. 1(d). Due to absorption of light, the resonator heats up, red-shifting the resonances at 1.3 μm and 1.5 μm. Note that the mode at 1.3 μm resonance shifts by a factor of 1.16=$\Delta f_{1330}/\Delta f_{1550}$ more than the 1.5 μm resonance as a result of the higher mode number. The 1.5 μm pump laser is now tuned into its resonance from the blue side, passing through a chaotic four-wave mixing regime. The rising intracavity pump power has the same thermal effect as the 1.3 μm auxiliary laser and also red-shifts both resonances (shown in Fig. 1(e)). This red shift of the resonances at 1.3 μm causes the 1.3 μm intracavity power to decrease, counteracting the temperature rise induced by the 1.5 μm pump laser. The temperature (and intracavity optical powers) of the microresonator will reach a stable equilibrium since the frequencies of both lasers are on the blue side of their optical resonance modes [42]. The intracavity power of the 1.3 μm auxiliary laser will be further reduced as the 1.5 μm pump laser scans into the resonance ($t_0$-$t_1$ in Fig. 1(g)). Once the pump laser gets close to zero detuning, the resonator abruptly transitions into the soliton regime. In this regime the pump resonance splits into a soliton resonance and a cavity resonance [20]. As shown in Fig. 1(f), the low-frequency and small peak is the soliton-induced 'S-resonance'. Upon entering the soliton regime, the intracavity pump power decreases abruptly, blue shifting the resonances. As a result, the intracavity auxiliary power passively rises to stabilize the temperature of the resonator ($t_1$...$t_2$ in Fig. 1(g)). Figure 1(h) shows experimental results when fixing the frequency of the 1.3 μm auxiliary laser on the blue detuned side of its resonance while simultaneously tuning the 1.5 μm pump laser into the pump resonance. We indeed see the 1.3 μm laser compensating the intracavity power variation, keeping the total circulating power inside of the resonator stable, therefore passively stabilizing the temperature of the microresonator. This effect can be optimized by varying the parameters (optical power and frequency detuning from the resonance) of the 1.3 μm laser, allowing the effective thermal response of silica resonators to be reduced by two orders of magnitude. An animation of the concept is available online [51].

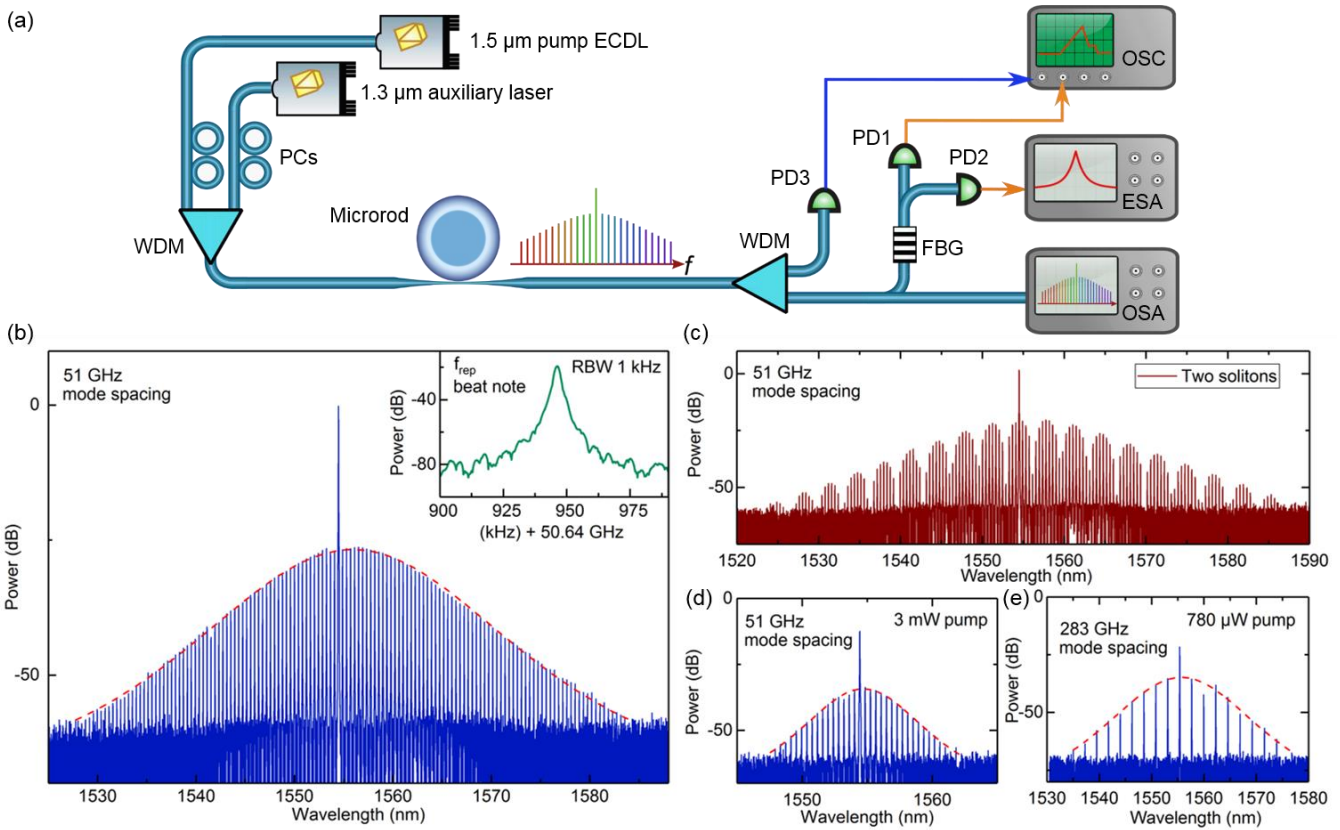

**Fig. 2.** (a) Experimental setup for the generation of a soliton mode-locked frequency comb in a silica microrod resonator by using an auxiliary laser to compensate the resonator's thermal shift. ECDL: external cavity diode laser; WDM: wavelength division multiplexer; PC: polarization controller; FBG: fiber Bragg grating; PD: photodetector; OSC: oscilloscope; OSA: optical spectrum analyzer; ESA: electronic spectrum analyzer. (b) Optical spectrum of a single-soliton state pumped with 80 mW optical power. The red dashed line shows the fitted $sech^2$ envelope. Inset: Spectrum of the microwave repetition rate $f_{rep}$ of the single-soliton state measured with a 1 kHz resolution bandwidth (RBW). (c) Optical spectrum of a two-soliton state pumped with 80 mW optical power. (d) Optical spectrum of a single soliton state pumped at 3 mW power with a fitted $sech^2$ envelope. (e) Optical spectrum of a single soliton state in a 230-μm-diameter microrod pumped with 780 μW power.

## 3. EXPERIMENTAL SETUP

Figure 2(a) shows the schematic of the experimental setup. A 1.5 μm external cavity diode laser (ECDL) with a short-term linewidth of <10 kHz is used as the pump laser for generating a soliton frequency comb. A 1.3 μm laser is used as an auxiliary laser to compensate changes of the circulating power in the resonator. As mentioned above, as well as choosing a high-Q mode family with minimal avoided crossings for soliton generation, we also choose a high-Q optical mode at 1334 nm, which enables us to operate the auxiliary laser at a low power level similar to the pump laser power. The two lasers are combined with a WDM and evanescently coupled to the microresonator via a tapered optical fiber. Two fiber polarization controllers (PCs) are used to optimize the coupling efficiency of the auxiliary and pump light into the microresonator. As shown in Fig. 1(c), a 1.3 mm diameter microrod resonator is used for soliton generation with a FSR of 50.6 GHz. The optical modes used to generate the soliton frequency comb have a quality factor of $3.7\times10^8$ with a ~520 kHz linewidth (measured at 1554 nm) while the chosen auxiliary optical mode has a quality factor of $2\times10^8$ with a ~1.1 MHz linewidth (measured at 1334 nm). At the resonator output, the auxiliary and pump lights are separated by another WDM. One part of the 1.5 μm light is sent to an optical spectrum analyzer (OSA), and the rest is sent into a fiber Bragg grating notch filter to separate the generated comb light from the pump light. The comb light is sent to two photodetectors: one for monitoring the comb power (PD1), and the other (PD2) with a 50 GHz bandwidth for detecting the repetition rate of the generated soliton frequency comb on an electronic spectrum analyzer (ESA). The auxiliary light is monitored by a third photodiode (PD3).

By optimizing the optical power of the 1334 nm auxiliary laser and its detuning from the optical resonance, both multi- and single- soliton states can be accessed by manually forward-tuning the pump laser into the soliton steps. Figure 2(b) shows the optical spectrum of a single soliton at a pump power of ~80 mW while ~60 mW of 1334 nm auxiliary power is used to compensate the thermal effect. The optical spectrum of the single soliton has a smooth, sech²-like shape (red dashed line in Fig. 2(b)). Note that there is no significant avoided-crossing behavior visible in the spectrum range. The 3 dB bandwidth of the spectrum is around 1.3 THz, corresponding to a 240 fs optical pulse. Once excited, the soliton states can survive for many hours without active feedback locking. For longer soliton lifetimes, active feedback locking can be used [44]. The soliton state is further confirmed by measuring the RF spectrum at the comb's 50.6 GHz repetition rate ($f_{rep}$). The beat note from the photodiode is amplified and mixed down to 13.6 GHz with a 37 GHz microwave signal from a signal generator. The down-converted spectrum at 13.6 GHz is analyzed with an electronic spectrum analyzer. The inset in Fig. 2(b) shows the $f_{rep}$ beat note when the microcomb is in the single-soliton state. Figure 2(c) shows the optical spectrum of a two-soliton state at a pump power of ~80 mW. The demonstrated technique constitutes a simple and robust way to access single solitons in microresonators making the system insensitive to pump laser frequency and power fluctuations. In particular this technique enables to access soliton states in resonators without mode crossings that naturally exhibit very narrow soliton steps.

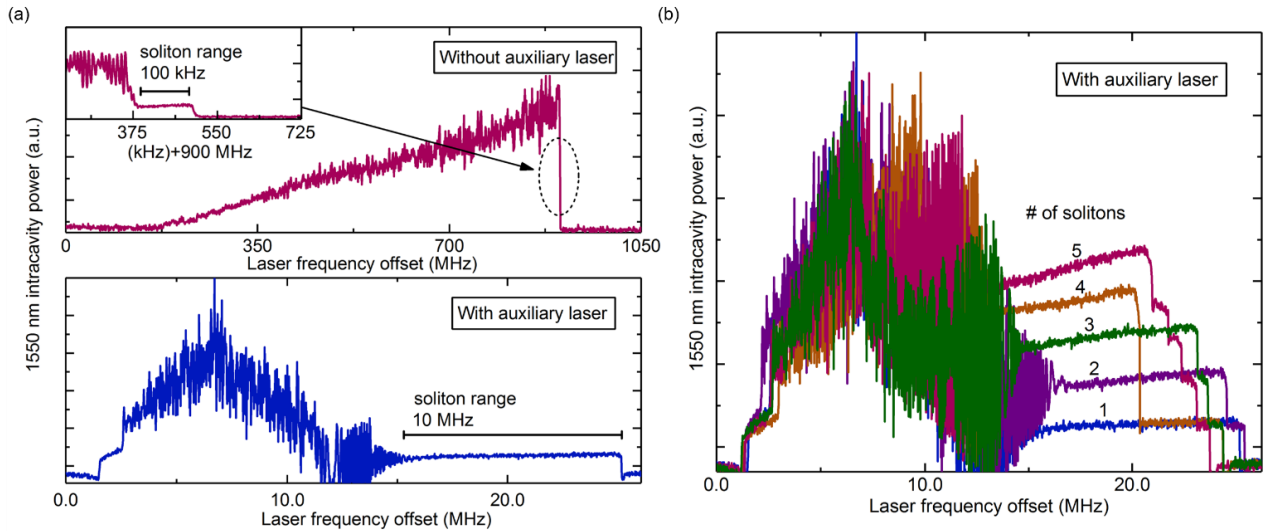

**Fig. 3.** Enhancement of soliton stability range. (a) Experimental traces of the 1550 nm intracavity power when scanning the pump laser frequency from blue to red detuning "without auxiliary laser" (upper panel) and "with auxiliary laser" (lower panel). The laser tuning speed is ~35 MHz/ms. The inset in the upper panel shows a trace of the narrow soliton step with a width of ~100 kHz without auxiliary laser. The lower panel shows the same resonance while the auxiliary laser is coupled into the resonator. The overall thermally broadened width of the resonance is reduced while the soliton stability range is increased by two orders of magnitude to ~10 MHz. (b) Experimental traces of the 1550 nm intracavity power for different multi-soliton states (with auxiliary laser).

To explore the full advantages of our proposed technique, we obtain a single-soliton frequency comb with both the pump laser and the auxiliary laser operating at very low optical powers. Figure 2(d) shows the optical spectrum of a single soliton state pumped by 3 mW optical power at 1554 nm with a few mW at 1334 nm auxiliary laser for compensating the thermal effect. The spectrum has a smooth sech²-like shape (red dashed line). In addition, using a smaller diameter (230 μm) microrod, a single-soliton state is accessed with 780 μW optical power (pump power in the tapered optical fiber), as shown in Fig. 2(e). To the best of our knowledge this is the first demonstration of a soliton microcomb at sub-mW power levels.

## 4. INCREASE OF THE SOLITON STABILITY RANGE

Soliton mode-locked states are generated by scanning the pump laser frequency from blue detuning to red detuning with respect to the resonator mode. Due to thermally induced (and Kerr effect induced) resonance frequency shifts, the measured power of the generated comb modes has a triangular shape [42]. This is shown in the upper panel of Fig. 3(a) for a measurement without the auxiliary laser. The pump wavelength, optical power, and laser scan speed are ~1554 nm, ~20 mW and ~35 MHz/ms, respectively. The width of the thermal triangle is ~700 MHz. At the end of the triangle shape (marked with a dash circle), "step-like" features are observable, which indicate the presence of soliton states. The inset in the upper panel of Fig. 3(a) shows a zoom into a single soliton step without auxiliary laser with a width of ~100 kHz (corresponding to a few microseconds for the used laser sweep speed). These kHz-scale soliton steps are usually too narrow to reliably generate solitons since any jitter of the pump laser frequency would lead to a loss of the thermally locked microresonator resonance.

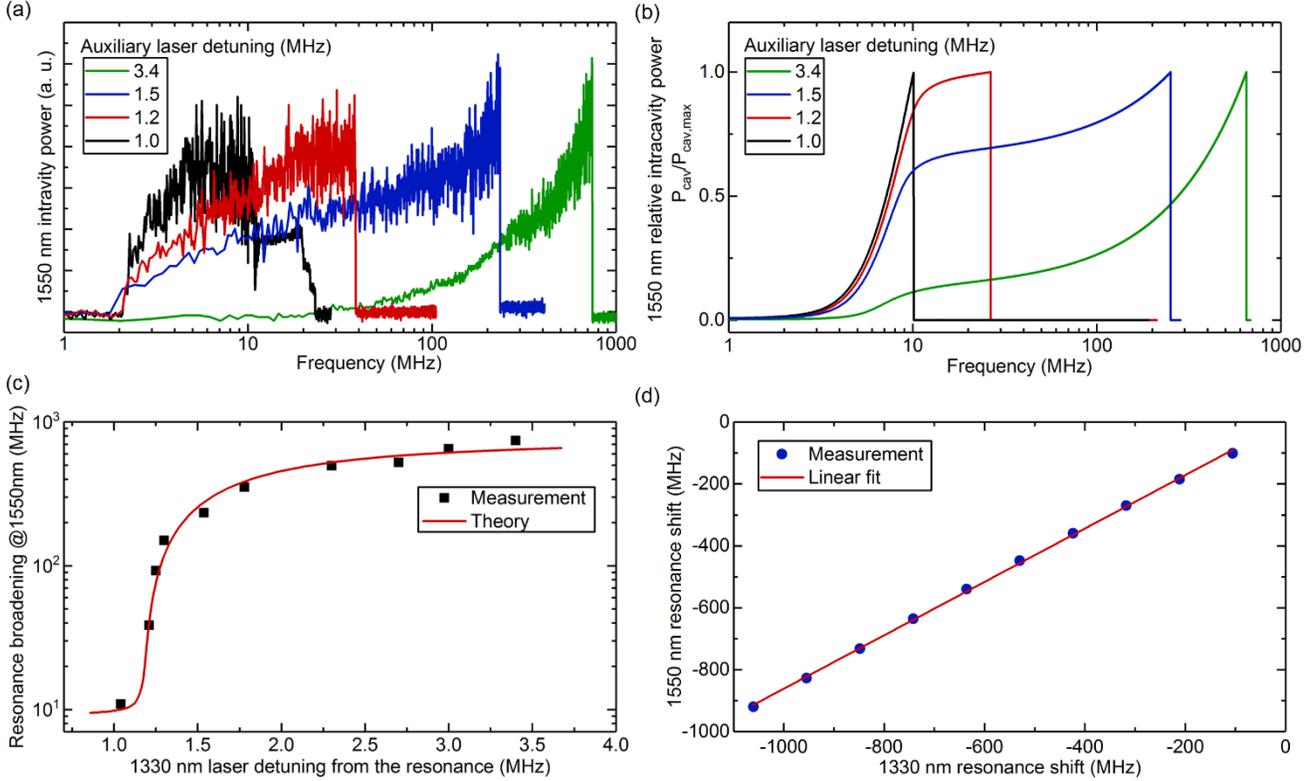

**Fig. 4.** Tuning behavior of the doubly pumped resonator system. (a) (b) Experimental and numerically calculated traces of the shape of the thermally broadened resonances at 1550 nm at different fixed detunings of the 1330 nm laser from the auxiliary resonance. The 1550 nm pump laser is tuned at ~35 MHz/ms during the measurement. (c) Experimental measurement (squares) of the width of 1550 nm resonances as a function of the detuning of the 1330 nm auxiliary laser from its resonance. The red line is a simulation based on the thermal dynamic behavior of a doubly pumped microresonator. (d) Experimental measurement and linear fit of the relative resonance shift of the 1550 nm and 1330 nm modes with a slope of 0.862.

In contrast, the lower panel in Fig. 3(a) shows the microresonator resonance when the 1334 nm auxiliary laser is simultaneously coupled into the resonator. The 1554 nm pump laser is operated at the same parameters (laser scan speed and optical power) as in the measurement without auxiliary laser. With the auxiliary laser coupled into the resonator, the length of the single-soliton step is significantly increased by two orders of magnitude to more than 10 MHz. Note that the overall width of the thermally broadened resonance is reduced to ~25 MHz, such that the soliton regime spans nearly half of resonance frequency range. Figure 3(b) shows a few examples of different soliton steps at 1554 nm when the auxiliary laser is close to its resonance. Each step represents a different integer number of solitons circulating inside the resonator. Note that, with the auxiliary laser, the length of soliton steps in time depends on the laser scanning speed, instead of the thermal relaxation velocity. As a result, when scanning more slowly (~1MHz/ms), the soliton steps can even last for tens of milliseconds, which is four orders longer than that without the auxiliary laser. Figure 4(a) shows the influence of the auxiliary laser detuning from its resonance on the thermally broadened width of the 1554 nm resonance. During this measurement, the auxiliary resonance thermally stabilizes itself to the auxiliary laser. It can be seen that smaller detuning of the auxiliary laser leads to a better stabilization of the circulating intracavity power and thus a reduced thermal broadening effect at the 1554-nm-resonance. A mathematical description of the narrowing can be found in the Supplementary Information. Figure 4(b) shows the numerical calculations, based on the dynamical thermal behavior of the microresonator resonance in the presence of an auxiliary laser in a second resonance. The results are consistent with the experimental measurements in Fig. 4(a). Figure 4(c) shows both the experimentally measured and the numerically calculated width of the 1554-nm-resonance as a function of the detuning of the 1334-nm-laser from the auxiliary resonance.

All the previous calculations take into account a larger thermally induced resonance frequency shift of the mode at 1334 nm as a result of the higher mode number. We verify this by measuring the relative resonance shift of pump resonance and auxiliary resonance when heating up the resonator with one of the lasers (see Supplementary Information for more details). The results are displayed in Fig. 4(d) and show a slope of $\Delta f_{1550}/\Delta f_{1330}$ = 0.862, which is close to the expected value of 0.859 based on the ratio of the mode numbers. This larger resonance shift at 1.3 μm gives the auxiliary laser more leverage on the resonator temperature and enables its operation at reduced power compared to the pump laser.

## 5. CONCLUSION

In summary, we have demonstrated that the stability of microresonator mode spectra can be greatly enhanced by coupling an auxiliary laser into a second high-Q resonance. This increase the laser frequency range in which a microresonator generates Kerr solitons by two orders of magnitude and significantly reduces the sensitivity of microcomb states to pump laser frequency and power fluctuations. The scheme enables long-term optical frequency comb generation without active stabilization of pump laser frequency or power. This greatly relaxes stability requirement for laser sources in future fully chip integrated microcomb systems. The enhanced stability enables us to demonstrate the first microresonator solitons at sub-mW power levels. In addition, the auxiliary laser could be used as an active actuator to stabilize a soliton frequency comb as shown in [52]. We believe that our technique of passively stabilizing microresonator mode spectra could be applied to other resonator systems/applications that require insensitivity to perturbations by an external laser.

**Funding**. H2020 Marie Sklodowska-Curie Actions (MSCA) (748519, CoLiDR); Horizon 2020 Marie Sklodowska-Curie grant (GA-2015-713694); National Physical Laboratory Strategic Research; H2020 European Research Council (ERC) Starting Grant (756966, CounterLight); LDB and MTMW acknowledge funding from the Engineering and Physical Sciences Research Council (EPSRC) via the CDT for Applied Photonics.